\documentclass[floatfix,twocolumn]{revtex4}

\usepackage{times,setspace}
\usepackage{amssymb,amsmath,graphicx}
\usepackage{graphicx}
\usepackage{amsmath,amssymb, amsthm}
\usepackage{rotating}
\usepackage{psfrag}
\usepackage[usenames]{color}

\newcommand{\bee}{\begin{equation}}
\newcommand{\ee}{\end{equation}}

\begin{document}
%
\title{{The effect of multiple paternity on genetic diversity during and after colonisation}}
\author{Marina Rafajlovi{\'c}$^{1}$, Anders Eriksson$^{2}$, Anna Rimark$^{1}$, Sara H. Saltin$^{3}$, Gregory Charrier$^{3}$, Marina Panova$^{3}$, Carl Andr\'e$^{3}$, Kerstin Johannesson$^{3}$ and Bernhard Mehlig$^{1}$\\
$^1$\emph{\small Department of Physics, University of Gothenburg, SE-41296 Gothenburg, Sweden}\\
$^2$\emph{\small Department of Zoology, University of Cambridge, CB2 3EJ Cambridge, UK}\\
$^3$\emph{\small Department of Biological and Environmental Sciences-Tj\"arn\"o,University of Gothenburg, SE-45296 Str\"omstad, Sweden}}
%

%
\begin{abstract}
In metapopulations, genetic variation of local populations is influenced by the genetic content of the founders, and of migrants following establishment.  We analyse the effect of multiple paternity on genetic diversity  using a model in which 
the highly promiscuous marine snail {\it Littorina saxatilis} 
expands from a mainland 
to colonise initially empty islands of an archipelago.
Migrant females carry a large number of eggs fertilised by $1-10$ mates. We quantify the genetic diversity of the population 
in terms of its heterozygosity: initially during the transient
colonisation process, and at long times when the population has reached an equilibrium state
with migration.
During colonisation,
multiple paternity increases the heterozygosity by $10-300\%$ in comparison with the case of single paternity.  The equilibrium state, by 
contrast, is less strongly affected: multiple paternity gives rise to $10-50\%$ higher heterozygosity compared with single paternity. 
Further we find that far from the mainland, new mutations spreading from the mainland cause  bursts of high genetic diversity separated by long periods of low diversity. 
This effect is boosted by multiple paternity. We conclude that multiple paternity facilitates colonisation and maintenance of small populations, whether or not this is the main cause for the evolution of extreme promiscuity in {\it Littorina saxatilis}.
\end{abstract}
\maketitle

\section{Introduction}

When new local populations are established in a metapopulation, genetic variation within the newly founded populations is initially governed by the genetic content of founders. At a later stage, during continued input of variation through migration, the genetic composition of migrants may potentially contribute new variation and hence counteract loss by drift and selection. In brooding and sexual species, empty sites are most likely colonised by single fertilised females that bring a brood of offspring along, while founders of virgin females, males, and juveniles fail to mate in an empty site \cite{Johannesson88}. In brooding species, female promiscuity (the propensity to mate multiple males) may influence the genetic variation carried by the founders, and, if so, will have consequences on the effective population size of the new population. 

Females mating multiple males have broods of offspring sired by more than one male, unless sperm competition or cryptic female choice prevent this. Female promiscuity, once believed to be rare in nature, is observed in a number of animal species including mammals, amphibians, fishes and invertebrates \cite{Birkhead, Jennions, Avise}. Genotyping offspring of species with promiscuous females shows that a large proportion of the females releases offspring sired by $2-4$ males \cite{Avise}. In some species of fish and invertebrates, levels of multiple paternity are even higher, since the number of males siring a brood regularly reaches $6-10$ \cite{Paterson,Trontti, Brante, Coleman}. An extreme example of high multiple paternity is the marine snail {\it Littorina saxatilis} with $15-23$ males siring broods of single females \cite{panova}. 
In this study we construct a mating model for this species and analyse how multiple paternity affects population genetic variation and structure in a metapopulation.
We consider established populations in equilibrium, but also populations under establishment (during initial colonisation of a previously empty habitat).

{\it Littorina saxatilis} is strictly intertidal and most abundant in rocky shores in the north Atlantic, with population densities of tens to hundreds of snails per square metre \cite{Reid}. In contrast to many other marine snails, {\it L. saxatilis} does not have pelagic eggs or larvae, and therefore dispersal over a few metres range is infrequent. However, snails occasionally migrate among islands, probably by rafting. It has been estimated that within an archipelago of small and large islands, $3\%$ of the small islands receive a migrant snail each year \cite{Johannesson95}. In many areas, {\it L. saxatilis} forms metapopulations with local populations inhabiting discrete localities, such as islands of an archipelago, rocky outcrops and breakwaters intermingled by sandy substrates \cite{Kerstin87,Johannesson90}. During the retreat of the ice sheet $12000-15000$ years ago, {\it L. saxatilis} spread from refuge areas both in the northern Atlantic and south of the ice-cap \cite{panova}. Part of this postglacial expansion comprised colonisation of hundreds of islands in the archipelago along the Swedish west coast that successively became available by isostatic uplift, a process that is still ongoing. In this system, populations on the mainland and large islands are the oldest and largest, and these are likely to act as the ultimate sources of genetic variation during colonisation of emerging islands in a stepwise manner (Fig.~\ref{fig:1}{\bf A}). We have re-analysed genetic data from {\it L. saxatilis} populations in the archipelago of west Sweden and found that the first principal component shows a largely linear relationship between population genetic variation and size/age of the islands/populations, with mainland populations at the one end, skerry populations (skerry sizes $\approx10 m^2$) at the other end, and island populations (island sizes $\approx 10^5 m^2$) in the middle (Fig.~\ref{fig:1}{\bf B}). This suggests a simple linear stepping-stone model with the mainland population acting as a source for colonisation of islands at successively younger age, and at increasing distance from the mainland (Fig.~\ref{fig:1}{\bf C}).

In this paper, we investigate how multiple paternity in {\it L. saxatilis} affects spatial and temporal structure of neutral genetic diversity in a metapopulation of this species. We analyse the effective genetic population size resulting from the mating behaviour observed in earlier studies, and derive simple approximations describing the genetic diversity of populations during colonisation, and in its equilibrium state with migration. We use simulations to assess the temporal effects of migration on genetic diversity as new mutations from the mainland spread to distant islands.

\section{Description of the model}\label{sec:methods}
 We construct a stepping-stone colonisation model with the following basic assumptions mimicking how {\it L. saxatilis} colonises the post-glacial archipelago of western Sweden. (1) Colonisation is frequent and rapid, as rafted fertilised females release a few hundred offspring already in the first generation \cite{Johannesson95}. (2) Small skerries are likely to be colonised within a few years after emergence, and hence all newly established populations are limited by the small size of the habitat, resulting in census sizes of $\approx10^2-10^3$ \cite{Johannesson95}. (3) Colonisation is likely to take place in a stepping-stone manner with smaller and more distantly related islands being colonised from their closest already colonised islands. For simplicity, we consider a system consisting of a mainland and of linearly arranged islands of equal carrying capacities (substantially  smaller than that of the mainland). 
\subsection{Stepping-stone colonisation model}\label{subsec:model}

\begin{figure}[tp]
\centerline{
\includegraphics[angle=0,width=8.4cm]{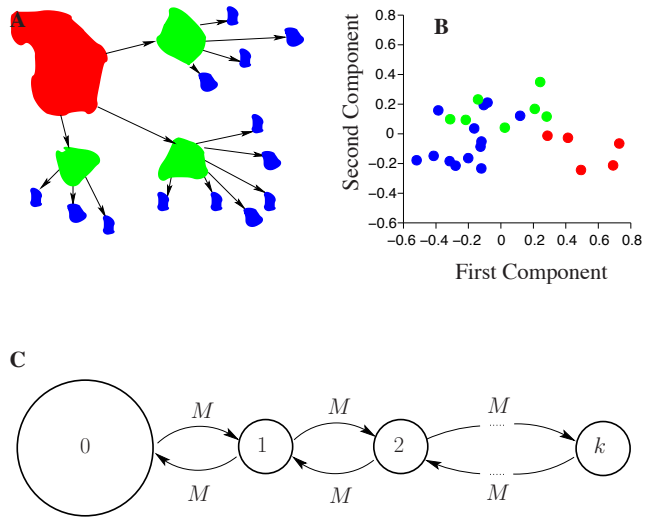}}\hspace*{1cm}
\caption{\label{fig:1} ({\bf A}) Physical structure of the marine habitats of {\it Littorina saxatilis}: mainland (red), islands (green), and skerries (blue). ({\bf B}) Principal components of allozyme population differentiation in {\it L. saxatilis} (data from \cite{Kerstin87}, the presumably selected locus $Aat-1$ is excluded). Populations are classified as mainlands (red), islands (green), or skerries (blue). ({\bf C}) Stepping-stone model of a section of the archipelago, with the mainland (labelled by $0$) acting as a source for establishing the island populations ($1$ to $k$).}
\end{figure}

In our model, islands are linearly arranged and numbered from $1$ to $k$, with $k$ being furthest away from the mainland (see Fig.~\ref{fig:1}{\bf C}). We include high values of $k$ in our model (such as $k=10$) in order to be able to assess saturation effects. The mainland is labelled by $0$. Generations are assumed to be non-overlapping. At the generation when the process of colonisation starts, the mainland is the only populated habitat, and the population heterozygosity on the mainland is stationary (that is, the mainland population is assumed to be old).

Within our model, an empty island becomes fully colonised in a single generation after the arrival of one or more founder females from the nearest neighbouring island. This is motivated by the very large capacity for population growth of {\it L. saxatilis} in a suitable habitat \cite{Kerstin87}. In our model the founder females give rise to $2N$ offspring in total, with equal sex ratio, where $2N$ denotes the carrying capacity of an island. Upon establishing a given island population, its population size remains constant over time. In our model, mating takes place before migration, which allows us to trace only the movement of females (males also migrate, but since they will not mate after migration, they do not contribute to the progeny on the island they migrated to). Individuals are equally likely to migrate to each of their closest neighbours (but the mainland and the last island have only a single neighbour, Fig.~\ref{fig:1}{\bf C}). On average, $M$ females migrate per generation from one island to a neighbouring island, except for empty islands that only receive migrants. 

In addition to the above, we assume that the population size on the mainland is much larger than the population size of a colonised island (unlike other models which assume that all habitats have the same carrying capacity, see, for example, \cite{Aus:97}). This allows us to treat the mainland as the only source of genetic variation to the island populations. In our computer simulations (see Sections~II-IV in the electronic supplementary material, ESM) we set the mainland heterozygosity  to unity. This simplifies the simulations, since the dynamics of the mainland does not need to be simulated explicitly. 

\subsection{Mating model}\label{subsec:mating}

In order to study the consequences of multiple paternity for genetic diversity, we introduce a mating model to describe different levels of multiple paternity in mating systems. 

Based on known life-history traits \cite{Kerstin87}, we assume that  the duration of the reproductive cycle of females is the same for all females. Each female carries beneath her shell juveniles of varying degrees of maturity, and juveniles are released at an approximately steady rate. We also assume that after a successful mating, the mated female obtains a sperm package able to fertilise female eggs during its persistence time. Our observations show that sperm can be stored and used up to a year after mating.
The number of eggs fertilised by a single sperm package is denoted by $N_{\rm eggs}$, and we assume that this number is the same for all sperm packages that a female receives during the reproductive cycle. The total number of sperm packages received by a female during her reproductive cycle  is denoted by $l$. 

The probability $p$ that two eggs are fertilised by the same sperm package is $p=l^{-1}$ assuming that all sperm packages a female received during her reproductive cycle persist until the end of the reproductive cycle, that all eggs are fertilised after all sperm packages have been collected, and that sperm packages are chosen with replacement to fertilise eggs. 

The scheme presented above models the process of mating at an individual level. We assume that individuals belong to a well mixed diploid population of $N_{\rm m}$ males and $N_{\rm f}$ females, and we take $N_{\rm m}\gg 1$ and $N_{\rm f}\gg 1$. In our model, a female encounters $s\leq N_{\rm m}$ different males  during her reproductive cycle. Having $s< N_{\rm m}$ reflects the limited movement of snails during the reproductive cycle. To simplify the analysis, we assume that all males a female encounters are equally likely to be her mating partner in any of the matings she experiences. Moreover, we assume that all females are equally successful mothers. Within the model described, the effective size of a single local population is given by (see Section~I of the ESM):
 
 \bee\label{eq:ne}
 N_{\rm e}=4\left(\frac{1+\kappa}{N_{\rm f}}+\frac{1}{N_{\rm m}}\right)^{-1}\,\,.
 \ee
Here, $\kappa$ is the probability that two offspring having the same mother share a father

\bee\label{eq:kappa}
 \kappa=p+\left(1-p\right)\frac{1}{s}\,\,.
\ee 
Since $\kappa$ decreases as the number of available mates $s$ of a female increases (keeping the value of $p<1$ fixed), we take $\kappa^{-1}$ as a measure of the degree of multiple paternity.

Our model reduces to the model in \cite{BalLeh} in the case a female encounters all males present in the population
(upon substituting the number of matings in \cite{BalLeh} by $p^{-1}$). If each female mates with all males in the population 
and the probability that two eggs are fertilised by the same sperm package is $p=0$, our model reduces to random mating.

\begin{figure}[tp]
\centerline{
\includegraphics[angle=0,width=5cm]{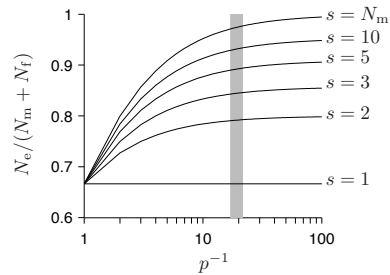}}\hspace*{1cm}
\caption{\label{fig:neff} Effective population size (for the given number of available mates, $s$, and the probability that two eggs are fertilised by the same sperm package, $p$) relative to $N_{\rm m}+N_{\rm f}$. The grey region depicts the average number of fathers in four broods of {\it L. saxatilis} \cite{panova}. Parameters: $N_{\rm m}=N_{\rm f}=10^3$.}
\end{figure}

We show in Fig.~\ref{fig:neff} that the effective population size increases as $s$ increases. For the parameters set in Fig.~\ref{fig:neff}, the maximum value of $N_{\rm e}$ is equal to $N_{\rm m}+ N_{\rm f}$, which corresponds to the effective population size under random mating. The increasing trend of $N_{\rm e}$ saturates at $s\approx 10$ for a given value of $p$. 
In summary, by increasing the degree of multiple paternity, the effective population size becomes larger (as found in \cite{Pearse} for a different mating model).

We compared the male family sizes (the number of offspring of a father involved in siring the brood of a given female) obtained under our model to those obtained under experimental conditions, as well as in natural populations. We conducted experiments such that $6$ virgin females were placed in separate aquaria, and each was accompanied with $s=10$ males. The sire of each offspring produced during a year was determined by genotyping 8 microsatellite loci.


\section{Results}\label{sec:gen_div}
\begin{figure}[tp]
\centerline{
\includegraphics[angle=0,width=8.4cm]{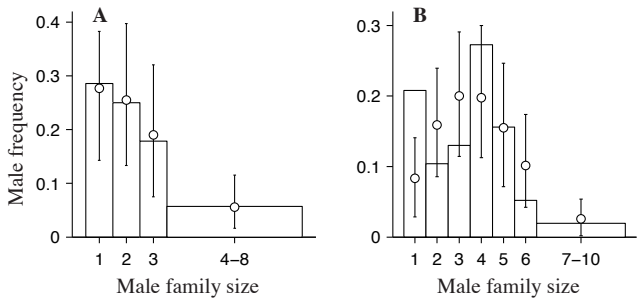}}\hspace*{1cm}
\caption{\label{fig:2} Histograms of male family sizes within broods of females. Bars in panel {\bf A} show the empirical data obtained under experimental conditions for $s=10$; the data correspond to six broods, two of size $20$, three of size $19$, and one of size $16$. Bars in panel {\bf B} show results taken from \cite{panova}; the data correspond to four broods of sizes $79$, $77$, $71$, and $53$. The width of the bins are chosen so that the expected number of counts in each bin is not smaller than $5$. The probability assigned to each bar is proportional to the bar area. Symbols and error bars show the result of the best fit to the experimental data, together with their 95\% confidence intervals: $p=1/15$ in {\bf A}, and $p=0,~s=20$ in {\bf B}. We simulated  $10^3$ independent realisations of the mating process.}
\end{figure}

 In Fig.~\ref{fig:2}{\bf A} we show  the histograms of male family sizes obtained experimentally and in  Fig.~\ref{fig:2}{\bf B} we show similar data collected from females in natural habitats \cite{panova}. For both sets of data we use computer simulations in order to find the parameters in our model resulting in male family sizes that are closest to those empirically observed (the brood sizes analysed in computer simulations correspond to those from the empirical data). For data obtained under experimental conditions, we vary the probability $p$ in the computer simulations, and  for data collected in natural habitats, we vary both $s$ and $p$. We use a $\chi^2$-test to quantify the distance between the empirical and simulated data. The best fits obtained are shown in Fig.~\ref{fig:2}{\bf A}-{\bf B} by circles, and they correspond to $p=1/15$ (Fig.~\ref{fig:2}{\bf A}), and to $s=20$, $p=0$ (Fig.~\ref{fig:2}{\bf B}). In summary, Fig.~\ref{fig:2} suggests that our mating model describes empirical data obtained in the experimental setup well, whereas the agreement between the model and the empirical data taken from natural populations is less good. This is discussed in Section~\ref{sec:conc}. 

To address the question of how multiple paternity affects genetic variation and structure in our metapopulation, we analyse genetic diversity under the colonisation model described in Methods.  We analyse separately two phases of population dynamics on each island: initial colonisation, and the equilibrium state  
that develops once the colonisation phase is over.
For a given island, we compute the expected heterozygosity in the generation when the island is colonised (colonisation phase), as well as the heterozygosity in the equilibrium state.  
The corresponding analytical computations are described in detail in 
Sections~II and III of the ESM. We also study the temporal changes of heterozygosity under our model by computer simulations. In the following two subsections we present separately the results for the colonisation phase and for the equilibrium state.  

\subsection{Colonisation phase}\label{sub:colonis} 

\begin{figure}[tp]
\centerline{
\includegraphics[angle=0,width=8.4cm]{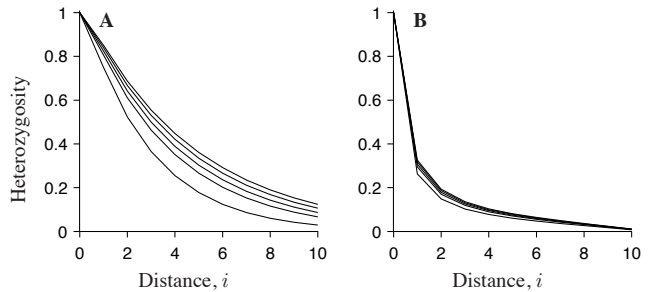}}\hspace*{1cm}
\caption{\label{fig:col_het} Analytically computed heterozygosity during colonisation ({\bf A}), and in the equilibrium state ({\bf B}). The lines shown from top to bottom correspond to: $s=10$, $s=5$, $s=3$, $s=2$, and $s=1$. Remaining parameters: the mainland heterozygosity is set to unity, scaled female migration rate $M=0.05$, number of females in each populated island $N=100$, probability that two eggs are fertilised by the same sperm package $p=0.1$, number of islands $k=10$.}
\end{figure}

We compute the heterozygosity during colonisation analytically using a coalescent approach \cite{Kingman:1982}. We represent the population-size history of the population on island $i$ as a sequence of $i$ bottlenecks ($i$ is the number of colonisation events that the population ancestral to that on island $i$ went through before the island was colonised). Our analytical result is valid for small migration rate, $M\ll 1$. Under this assumption, colonisation of empty islands occurs rarely, but when it does, an island typically receives a single founder female (see Section~II of the ESM). The result is given in Eq.~(14) of the ESM and in Fig.~\ref{fig:col_het}{\bf A}. We see that the colonisation-phase heterozygosity decays as distance from the mainland increases. We note that the results of our computer simulations  (see Fig.~1{\bf A}-{\bf C} in the ESM) agree well with the analytical results for low migration rates. For large migration rates ($M=0.5$. i\,.e\,. $0.5$ females on average per generation) by contrast, the simulations assume somewhat higher values than the theory. The reason for this deviation is that  for $M=0.5$ it is probable that more than one founder female comes to an empty island to establish the population, and, consequently, will contribute with more genetic variation than just one founder female. For natural populations of {\it L. saxatilis} it has been estimated that $3\%$ of empty islands receive a migrant each year  \cite{Johannesson95}. This estimate is close to the lower value of $M$ used in our simulations ($M=0.05$). An important result shown in Fig.~\ref{fig:col_het}{\bf A} is that at any distance from the mainland, multiple paternity results in higher heterozygosity than single paternity. Mating two males  ($s=2$) increases the values of single-paternity heterozygosity by $10-100\%$ for the parameters used in Fig.~\ref{fig:col_het}{\bf A}, and mating ten males ($s=10$) increases the values of single-paternity heterozygosity by $10-300\%$ (Fig.~\ref{fig:col_het}{\bf A}). The largest increase is observed at the island furthest from the mainland.  We also note that mating more than about $10$ males only marginally increases the heterozygosity (results not shown), as found in the case of freely mixing populations (see Subsection~\ref{subsec:mating}).

\subsection{Equilibrium state}\label{sub:long_times}
We show in Section~III of the ESM how to compute the heterozygosity in the equilibrium state at distance $i$ from the mainland using recursion relations (see, for example, \cite{Wright:1931}). Note that this derivation does not require $M$ to be small.

The results are given in Section III of the ESM and in Fig.~\ref{fig:col_het}{\bf B}. As in the colonisation phase, the equilibrium-state heterozygosity decreases as distance from the mainland increases. 
Also, by increasing the degree of multiple paternity, the heterozygosity at a given island increases (this effect saturates at $s\approx 10$, results not shown).  In contrast to the strong effect of multiple paternity during colonisation, the effect is substantially smaller in the equilibrium state. We find that the single-paternity heterozygosity in the equilibrium state increases by $10-20\%$ for $s=2$, and by $20-50\%$ for $s=10$ (Fig.~\ref{fig:col_het}{\bf B}). As in the colonisation phase, the largest increase is observed at the island furthest from the mainland.

\begin{figure}[tp]
\centerline{
\includegraphics[angle=0,width=9cm]{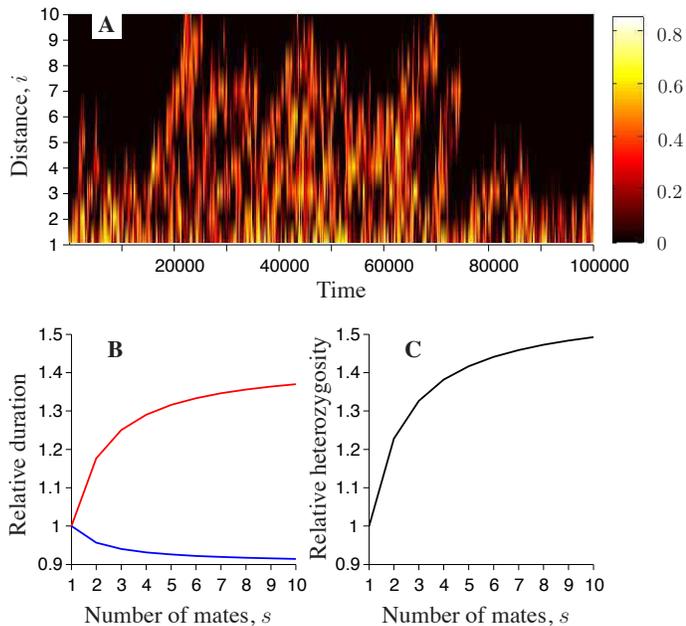}}\hspace*{1cm}
\caption{\label{fig:merged2} ({\bf A}) Heterozygosity as a function of the distance from the mainland and of time (single realisation of the model described). Mainland is not shown. The data correspond to $10^5$ generations after the initial $7\cdot10^6$ generations. The number of available mates is $s=10$. ({\bf B}) Durations of low-and high-heterozygosity phases (blue, and red) relative to their corresponding values for $s=1$. {\bf C} Equilibrium-state heterozygosity (black) relative to its corresponding value for $s=1$. Remaining parameters are as in Fig.~\ref{fig:col_het}{\bf B}.}
\end{figure}

In addition, we examined the variation in heterozygosity over consecutive generations in a particular realisation of our model. We find that the heterozygosity shows strong temporal fluctuations (Fig.~\ref{fig:merged2}{\bf A}). Notably, the fluctuations are strongest furthest from the mainland, with periods of high diversity separated by long periods of near or complete fixation. Hence the distribution of heterozygosity at large distance from the mainland is bimodal. The heterozygosity is expected to have a bimodal distribution in the case of a very small rate of income of new genetic material, as pointed out in \cite{Mehlig_het} (see Fig.~1 in \cite{Mehlig_het}).

In what follows, we analyse how the durations of the phases of low and high heterozygosity are affected by multiple paternity. Using the results of computer simulations, we compute the average durations of low- and high-heterozygosity phases at the island furthest from the mainland. We also derive  corresponding analytical results under the assumption that the scaled migration rate $M$ is small (see  
Section~IV of the ESM). For island $i=10$ we show  in Fig.~\ref{fig:merged2}{\bf B} the durations of low- and high-heterozygosity phases relative to their corresponding values for a single mate ($s=1$).                                
Fig.~\ref{fig:merged2}{\bf B}  shows that multiple paternity prolongs the duration of the high-heterozygosity phase, and decreases the duration of the low-heterozygosity phase. For example, the high-heterozygosity phase for the highest level of multiple paternity shown ($s=10$) is prolonged by around $40\%$ compared to its value under single paternity  ($s=1$). The low-heterozygosity phase is shortened by only around $10\%$ for $s=10$ (Fig.~\ref{fig:merged2}{\bf B}). For comparison, Fig.~\ref{fig:merged2}{\bf C} shows the equilibrium-state heterozygosity relative to its corresponding value for a single mate ($s=1$).  In conclusion, multiple paternity promotes heterozygosity by prolonging the duration of peaks of variation that reach the most distant islands.

\section{Discussion and conclusions}\label{sec:conc}

In this study we analysed the effect of multiple paternity on genetic diversity over spatial and temporal scales in a metapopulation. We quantified the effect of multiple paternity during the colonisation of semi-isolated populations and in the equilibrium state developed after the colonisation phase. Our conclusions given below can be  generalised to a metapopulation of patches that are partly isolated from each other, for example, by sandy beaches, harbors or other types of less suitable habitats.

We introduced a mating model which allows for different levels of multiple paternity in a population. In order to determine how realistic our mating model is, we compared the male family sizes of female broods obtained within our model, to empirically observed family sizes in populations of {\it L. saxatilis} from natural habitats and under experimental conditions. We found that male family size distributions from the experiments, where the true number of mates was known, are in very good agreement with our mating model.
The best-fitting parameters indicate fewer matings than the brood size, suggesting that some of the eggs are fertilised by sperm packages retained between matings.
By contrast, the corresponding empirical distribution in natural populations is best fitted by assuming an unlimited number of matings (i.e. no sperm retention). This is consistent with the high density of snails observed in the wild. 
However, the empirical distribution shows an excess of males with a single progeny compared to the mating model with the best-fitting parameters. This discrepancy could be explained by post-zygotic selection, where sperm from several males compete, resulting in uneven success among males. 
However, this effect should also be present in the experiments, but we did not find any evidence of it in the data. Another possibility is that this pattern is due to variation in individual snails movements in the wild, where some
snails might move around more extensively and mate with a new partner each time
while others stay within a small area and mostly re-mate with
individuals in the close surroundings. This variation in mating behavior cannot happen in the experiments where the snails are confined to each other.

Within our mating model,  by increasing the degree of multiple paternity, the effective population size increases, and thus the heterozygosity increases. However, since mating is considered to be costly \cite{panova}, it is not clear whether or not mating multiple males is an evolutionary strategy to increase the heterozygosity. Recall that we estimated that in natural populations of {\it L. saxatilis} the probability that two eggs are fertilised by the same sperm package is likely to be very small. Under our model, this probability is equal to zero if each sperm package fertilises one egg, or if the actual number of sperm packages a female receives during her reproductive cycle is very large. If the latter applies, we find that it is unlikely that the heterozygosity increase is the main reason for the extreme multiple paternity in this species. As earlier suggested, it seems likely that the cost of rejecting an intercourse is higher than the cost of accepting it, and a consequence of convenience polyandry \cite{panova}. Nevertheless, the consequences of multiple paternity for heterozygosity in relatively small and semi-isolated populations are substantial. This is summarised in the following.
 
At a given distance from the mainland, populations with high degree of multiple paternity establish higher heterozygosity than populations with low degree of multiple paternity. While this effect is substantial during colonisation, it is modest in the equilibrium state. This is explained in the following. Upon the arrival of founder females to an empty island, the carrying capacity of an island is reached within a single generation. Therefore, such a newly established population receives genetic material of most males that the founder females were inseminated by. By contrast, in the equilibrium state, all mothers present in an island contribute to the population in the next generation, and hence the impact of immigrant females to the next generation is rather small. From this reasoning, we find that it is possible that  the effect of multiple paternity 
upon heterozygosity during colonisation might decrease if the growth rate of the island populations up to the carrying capacity were less than infinite (as assumed in our model). 
 
The heterozygosity at distances far from the mainland fluctuates significantly. Long periods of almost complete loss of genetic variation are interrupted by bursts of high heterozygosity, and this effect is boosted by multiple paternity. The durations of high- and low-heterozygosity phases could be an important survival factor in natural populations. For example, the low-heterozygosity phase could be disadvantageous if a malignant disease appears in the population, assuming that only a particular mutation (not present in the population, or being rare) can survive the disease. 

The wave-like nature of the spread of new alleles from the mainland population is also seen in the  correlation of genetic diversity at neighbouring islands. We find that the correlation between heterozygosities at a pair of nearest-neighbouring islands increases as distance from the mainland increases  (results not shown). These results suggest intermittent bursts of genetic diversity in remote islands, an effect which becomes stronger  as the degree of multiple paternity increases. 

The conclusions given above are confirmed by our computer simulations. In order to minimise computing time during simulations, we assumed that the mainland heterozygosity is equal to unity, which guarantees that whenever a migrant from the mainland comes to the first island, it carries genetic material that previously existed neither on the mainland nor on any of islands (and thus the population dynamics on the mainland does not need to be simulated explicitly).  However, we emphasise that the conclusions given above qualitatively do not depend on the value of heterozygosity on the mainland.  
 
 We note that, unlike in our model, it is possible that the rate of successful colonisation in natural habitats is smaller than the rate of migration. For example, if an immigrant female carries a small number of progeny, her progeny alone may not be enough to colonise an empty island successfully. By allowing for the rate of successful colonisation to be smaller than the rate of migration in the colonisation model, the equilibrium-state values of heterozygosity remain equal to those obtained under the assumption that the colonisation and migration rates are the same (as in our model).  The values of heterozygosity during colonisation, by contrast, 
are expected to differ from those found here. 

In summary, this study can be used to quantify the gene flow between partly isolated natural populations using allelic frequencies at a number of neutral loci. Since our results show that the heterozygosity exhibits extreme fluctuations in populations founded through repeated founder events, we raise the question of whether similar fluctuations can be observed at any given time at neutral loci sampled genome-wide. 
In order to answer this and related questions, the effect of recombination needs to be analysed. 
Since island populations in our model  experience at least one severe bottleneck, we expect that the degree of linkage disequilibrium in the colonisation phase is constant over a range of genetic distances, as shown in \cite{Sch:12}. However, how multiple paternity affects linkage disequilibrium during colonisation and in equilibrium is yet to be understood. It would also be interesting to analyse how selection combined with recombination affects genetic diversity in a metapopulation. Such results would provide an advance in the endeavor of identifying genes under selection, and especially, the genes underlying speciation \cite{Nosil:2008,Johannesson:2010,Butlin}.


\textit{Acknowledgements.} We gratefully acknowledge financial support by Vetenskapsr\aa det, by the G\"oran Gustafsson Foundation for Research in Natural Sciences and Medicine, and by the platform ``Centre for Theoretical Biology'' at the University of Gothenburg. Further support was given from the Linnaeus Centre for Marine Evolutionary Biology (CeMEB, www.cemeb.science.gu.se).


\begin{thebibliography}{23}
 \bibitem{Johannesson88}
 Johannesson, K., 1988 The paradox of Rockall: why is a brooding gastropod ({\it Littorina saxatilis}) more widespread than one having a planktonic larval dispersal stage ({\it L. littorea})? {\it Mar. Biol.} {\bf 99}, 507--513.
 \bibitem{Birkhead}
 Birkhead, T. R., 2000 {\it Promiscuity: an evolutionary history of sperm competition.} Cambridge (MA): Harvard University Press.
 \bibitem{Jennions}
 Jennions, M. D. \& Petrie, M., 2000 Why do females mate multiply? A review of the genetic benefits. {\it Biol. Rev. Camb. Philos. Soc.} {\bf 75}, 21--64.
 \bibitem{Avise}
 Avise, J. C., Tatarenkov, A. \& Liu, J. X., 2011 Multiple mating and clutch size in invertebrate brooders versus pregnant vertebrates. {\it Proceedings of the National Academy of Sciences USA} {\bf 108}, 11512--11517.
 \bibitem{Paterson}
 Paterson, I. G., Partridge, V. \& Buckland-Nicks, J., 2001 Multiple paternity in {\it Littorina obtusata} (Gastropoda, Littorinidae) revealed by microsatellite analyses. {\it The Biological Bulletin} {\bf 200}, 261--267.
 \bibitem{Trontti}
 Trontti, K., Thurin, N., Sundstr\"om, L. \& Aron, S., 2007 Mating for convenience or genetic diversity? Mating patterns in the polygynous ant {\it Plagiolepis pygmaea}. {\it Behavioral Ecology} {\bf 18}, 298--303.
\bibitem{Brante}
Brante, A., Fernandez, M. \& Viard, F., 2011 Microsatellite evidence for sperm storage and multiple paternity in the marine gastropod {\it Crepiduda coquimbensis}. {\it J. Exp. Mar. Biol. Ecol.} {\bf 396}, 83--88.
\bibitem{Coleman}
Coleman, S. W. \& Jones, A. G., 2011 Patterns of multiple paternity and maternity in fishes. {\it Biological Journal of the Linnean Society} {\bf 103}, 735--760.
\bibitem{panova}
Panova, M., Bostr\"om, J., Hofving, T., Areskoug, T., Eriksson, A., Mehlig, B., M\"akinen, T., Andr\'e, C. \& Johannesson, K., 2010 Extreme female promiscuity in a non-social invertebrate species. {\it PLoS ONE} {\bf 5}, e9640.
\bibitem{Reid}
Reid, D. G., 1996 {\it Systematics and Evolution of Littorina}. London: The Ray Society.
\bibitem{Johannesson95}
Johannesson, K. \& Johannesson, B., 1995 Dispersal and population expansion in a direct developing marine snail ({\it Littorina saxatilis}) following a severe population bottleneck. {\it Hydrobiologia} {\bf 309}, 173--180.
\bibitem{Kerstin87}
Janson, K., 1987 Genetic drift in small and recently founded populations of the marine snail {\it Littorina saxatilis}. {\it Heredity} {\bf 58}, 31--37.
\bibitem{Johannesson90}
Johannesson, K. \& Warmoes, T., 1990 Rapid colonization of Belgian breakwaters by the direct developer {\it Littorina saxatilis} Olivi. {\it Hydrobiologia} {\bf 193}, 99--108.
\bibitem{Aus:97}
Austerlitz, F., JungMuller, B., Godelle, B. \& Gouyon, P., 1997 Evolution of coalescence times, genetic diversity and structure during colonization. {\it Theoretical Population Biology} {\bf 51}, 148Ð 164.
\bibitem{BalLeh}
Balloux, F. \& Lehmann, L., 2003 Random mating with a finite number of matings. {\it Genetics} {\bf 165}, 2313--2315.
\bibitem{Pearse}
Pearse, D. E. \& Anderson, E. C., 2009 Multiple paternity increases effective population size. {\it Molecular Ecology} {\bf 18}, 3124--3127.
\bibitem{Kingman:1982}
Kingman, J., 1982 The coalescent. {\it Stoch. Proc. Appl.} {\bf 13}, 235--248.
\bibitem{Wright:1931}
Wright, S., 1931 Evolution in Mendelian populations. {\it Genetics} {\bf 16}, 97Ð159.
\bibitem{Mehlig_het}
Eriksson, A., Haubold, B. \& Mehlig, B., 2002 Statistics of selectively neutral genetic variation. {\it Phys. Rev. E} {\bf 65}, 040901.
\bibitem{Sch:12}
Schaper, E., Eriksson, A., Rafajlovic, M., Sagitov, S. \& Mehlig, B., 2012 Linkage disequilibrium under recurrent bottlenecks. {\it Genetics} {\bf 190}, 217--229.
\bibitem{Nosil:2008}
Nosil, P., 2008 Speciation with gene flow could be common. {\it Molecular Ecology} {\bf 17}, 2103--2106.
\bibitem{Johannesson:2010}
Johannesson, K., Panova, M., Kemppainen, P., Andr\'e, C., Rol\'an-Alvarez, E. \& Butlin, R. K., 2010 Repeated evolution of reproductive isolation in a marine snail: unveiling mechanisms of speciation. {\it Philosophical Transactions of the Royal Society B: Biological Sciences} {\bf 365}, 1735--1747.
\bibitem{Butlin}
Butlin, R., 2012 What do we need to know about speciation? {\it Trends Ecol. Evol.} {\bf 27}, 27--39.
\end{thebibliography}
\end{document}


%
\title{{Electronic supplementary material for \lq The effect of multiple paternity on genetic diversity during and after colonisation\rq}}
\author{Marina Rafajlovi{\'c}$^{1}$, Anders Eriksson$^{2}$, Anna Rimark$^{1}$, Sara H. Saltin$^{3}$, Gregory Charrier$^{3}$, Marina Panova$^{3}$, Carl Andr\'e$^{3}$, Kerstin Johannesson$^{3}$ and Bernhard Mehlig$^{1}$\\
$^1$\emph{\small Department of Physics, University of Gothenburg, SE-41296 Gothenburg, Sweden}\\
$^2$\emph{\small Department of Zoology, University of Cambridge, CB2 3EJ Cambridge, UK}\\
$^3$\emph{\small Department of Biological and Environmental Sciences-Tj\"arn\"o,University of Gothenburg, SE-45296 Str\"omstad, Sweden}}
\maketitle

\section{Effective population size under multiple paternity}\label{sec:eff_size}
In this section we compute the effective population size under the mating model introduced in 
Section~II of the main text. We assume that a population is diploid, isolated, well-mixed and that it consists of  $N_{\rm f}$ females, and $N_{\rm m}$ males. We assume $N_{\rm f}\gg 1$ and $N_{\rm m}\gg 1$. 
Note that in a diploid population with effective population size $N_{\rm e}\gg 1$, the population homozygosity $F_\tau$ in generation $\tau$ is given by:

\begin{equation}\label{eq:Fg}
F_\tau\approx\frac{1}{2N_{\rm e}}\epsilon_\tau+(1-\frac{1}{2N_{\rm e}})\chi_\tau\,\,.
\end{equation} 
Here $\epsilon _\tau$ is the inbreeding coefficient which stands for the probability that two alleles within a single randomly chosen individual are identical at time $\tau$. The second term, $\chi_\tau$, is the coancestry, that is the probability that two alleles sampled in generation $\tau$ from two different randomly chosen individuals are identical. In what follows, we compute $F_\tau$ under our mating model, and thereafter we use Eq.~(\ref{eq:Fg}) to determine the corresponding effective population size. We assume that  mutation rate per generation per allele is $\mu\ll 1$, and we employ the infinite-alleles model  \cite{Kimura:64}. Our calculations are based on the approach used in \cite{BalLeh}.

 Under the assumption that all females are equally successful in producing offspring, it follows that the probability for two offspring to come from the same female, $P_{\rm f}$, is equal to $P_{\rm f}=N_{\rm f}^{-1}$. Let $\kappa$ be the probability that two offspring coming from the same mother also share a father. The probability that two offspring come from the same male, $P_{\rm m}$, is thus: 

\bee\label{eq:pm}
 P_{\rm m}=P_{\rm f}\kappa+(1-P_{\rm f})\frac{1}{N_{\rm m}}\,\,,
 \ee
where  $N_{\rm m}^{-1}$ is the probability that two offspring having different mothers stem from the same father. The probability $\kappa$ has two contributions: the probability that two offspring come from the same sperm package, $p$, and the probability that two offspring do not come from the same sperm package but they come from the packages coming from the same male, $(1-p)/s$. Here, we assume that sperm packages are chosen with replacement to fertilise eggs. It follows that $\kappa$ is given by
 
 \bee\label{eq:kap}
 \kappa=p+\left(1-p\right)\frac{1}{s}\,\,,
 \ee 
where $p$ is  

\begin{equation}\label{eq:p}
p=\frac{l\binom{N_{\rm eggs}}{2}}{\binom{l N_{\rm eggs}}{2}}\,\,.
\end{equation}  
As mentioned in Section~II of the main text, we take $\kappa^{-1}$ as a measure of the degree of multiple paternity.  The case of  single paternity corresponds to $\kappa=1$.

Using the expressions for $P_{\rm m}$ and $P_{\rm f}$, we compute   $\epsilon_\tau$, and $\chi_\tau$ recursively. Under our model we find

\begin{align}
\epsilon_\tau&=(1-\mu)^2\chi_{\tau-1}\,\,,\label{eq:ep0}\\
\chi_\tau&=(1-\mu)^2\Biggl\{\frac{1}{4N_{\rm f}}\left[\frac{1+\epsilon_{\tau-1}}{2}(1+\kappa)+\left(3-\kappa\right)\chi_{\tau-1}\right]\nonumber\\
&+\frac{1}{4}\left(1-\frac{1}{N_{\rm f}}\right)\left[\frac{1+\epsilon_{\tau-1}}{2N_{\rm m}}+\left(4-\frac{1}{N_{\rm m}}\right)\chi_{\tau-1}\right]\Biggr\}\,\,.\label{eq:chi0}
\end{align} 
After rearranging the terms in Eqs.~(\ref{eq:ep0})-(\ref{eq:chi0}), and by keeping only the leading-order terms, we obtain:

\begin{align}
\epsilon_\tau&=(1-2\mu)\chi_{\tau-1}\,\,,\label{eq:ep1}\\
\chi_\tau&=\frac{1}{8}\left(\frac{1+\kappa}{N_{\rm f}}+\frac{1}{N_{\rm m}}\right)+\frac{1}{8}\left(\frac{1+\kappa}{N_{\rm f}}+\frac{1}{N_{\rm m}}\right)\epsilon_{\tau-1}\nonumber\\
&+\Bigg[1-\frac{2}{8}\left(\frac{1+\kappa}{N_{\rm f}}+\frac{1}{N_{\rm m}}\right)\Bigg]\chi_{\tau-1} 
-2\mu\chi_{\tau-1}\,\,.\label{eq:ch1}
\end{align}
Using Eqs.~(\ref{eq:ep1})-(\ref{eq:ch1}) we find the standard expression for the equilibrium-state homozygosity 

 \begin{equation}\label{eq:F}
F=\frac{1}{1+\theta}\,\,,
\end{equation} 
where, as usual, $\theta=4\mu N_{\rm e}$ is the scaled mutation rate. The effective population size $N_{\rm e}$ is given by

\begin{equation}\label{eq:nef}
N_{\rm e}=4\left(\frac{1+\kappa}{N_{\rm f}}+\frac{1}{N_{\rm m}}\right)^{-1}\,\,.
\end{equation}
 In the case $N_{\rm m}=N_{\rm f}=N$ (as assumed for island populations in our colonisation model), Eq.~(\ref{eq:nef}) becomes:

\begin{equation}\label{eq:Ne}
N_{\rm e}=\frac{4N}{2+\kappa}\,\,.
\end{equation}
Upon setting $\kappa=0$, Eq.~(\ref{eq:Ne}) reduces to $N_{\rm e}=2N$, that is the effective population size becomes equal to the census population size. This is the maximum value that $N_{\rm e}$ can acquire in an isolated population of census size $N_{\rm m}+N_{\rm f}$.
\section{Heterozygosity in the colonisation phase}\label{sec:colon_phase}

In this section we compute the population heterozygosity in the colonisation phase. As mentioned in 
Section~II in the main text, it is assumed that only the mainland is  populated initially, and that its population  size is constant in time. Moreover, we assume that the mainland population exists for long time before the start of colonisation (which occurs at generation $\tau=0$). We denote the colonisation-phase heterozygosity on the mainland by $H_{\rm c}^{(0)}$.

In order to find an expression for  the heterozygosity of island $i$ in the generation when it is colonised, $H_{\rm c}^{(i)}$, we use the coalescent approach. Recall that a populated island is assumed to  consist of $N$ males and $N$ females ($N$ is large), and it is assumed that this population size is much smaller than that of the mainland. Moreover, in the following we assume that the migration rate $M$ is small, $M\ll 1$. The average time between two successive founder events is thus long, and typically one founder female arrives at an empty island (the probability that two females come simultaneously is of order $M^2$, and it is negligible for $M\ll 1$). Under this assumption,  the ancestral population size of the newly established population at island $i$ can be represented by a sequence of $i$ bottlenecks, such that each bottleneck lasts for one generation (since the founder female gives rise to $2N$ offspring), and the time between two successive bottlenecks is on average  $M^{-1}$ generations long. Upon expressing   the generation index $\tau$ by $t$ such that   $\tau=\lfloor 2 t N_{{\rm e}}\rfloor$, where $N_{\rm e}$ is given by Eq.~(\ref{eq:Ne}), the waiting time between two successive founder events is approximately  exponentially distributed with mean 

\begin{equation}
(2MN_{\rm e})^{-1}\,\,. 
\end{equation}
In order to compute the heterozygosity, we note that in our model, the mainland acts as the only source of genetic variation.  This allows us to argue the following. First, if the MRCA of two alleles  sampled randomly from the newly established population at island $i$ was born on island $j< i$ ($j\neq0$), the two alleles sampled are identical. Second, if the MRCA was born on the mainland, the two alleles are expected to be identical with probability $F_{\rm c}^{(0)}=1-H_{\rm c}^{(0)}$. Therefore, in order to compute $H_{\rm c}^{(i)}$, it suffices to determine the probability  that the MRCA of two lines sampled from the newly established population  at island $i$ stem from an allele that was born on the mainland, $P(0|i)$.

The probability $P(0|i)$ has two contributions. The first contribution is the probability that the MRCA of two alleles sampled at island $i$ is  not found during a bottleneck. We find this to be equal to $1-\frac{1}{8}(1+\kappa)$. The second contribution is the probability that the MRCA of two alleles is not found between two successive bottlenecks. This term is equal to  $2MN_{\rm e}(2MN_{\rm e}+1)^{-1}$. It follows that  $P(0|i)$ is given by

\begin{equation}
P(0|i)=\left(1-\frac{1}{8}(1+\kappa)\right)^i\left(\frac{2MN_{\rm e}}{2MN_{\rm e}+1}\right)^{i-1}\,\,.
\end{equation} 
Therefore, the colonisation-phase heterozygosity at island $i$ is:

\begin{equation}\label{eq:h_in}
H_{\rm c}^{(i)}=P(0|i)H_{\rm c}^{(0)}\,\,.
\end{equation}
Note that for the case described here, the population size switches between $2N$ ($N$ males and $N$ females during the waiting time before the colonisation of the next island) and unity (one inseminated mother during a bottleneck). Therefore, the probability $\kappa$ does not enter Eq.~(\ref{eq:h_in}) only through the expression for $N_{\rm e}$, Eq.~(\ref{eq:Ne}).

\begin{figure}[tp]
\centerline{
\includegraphics[angle=0,width=13cm]{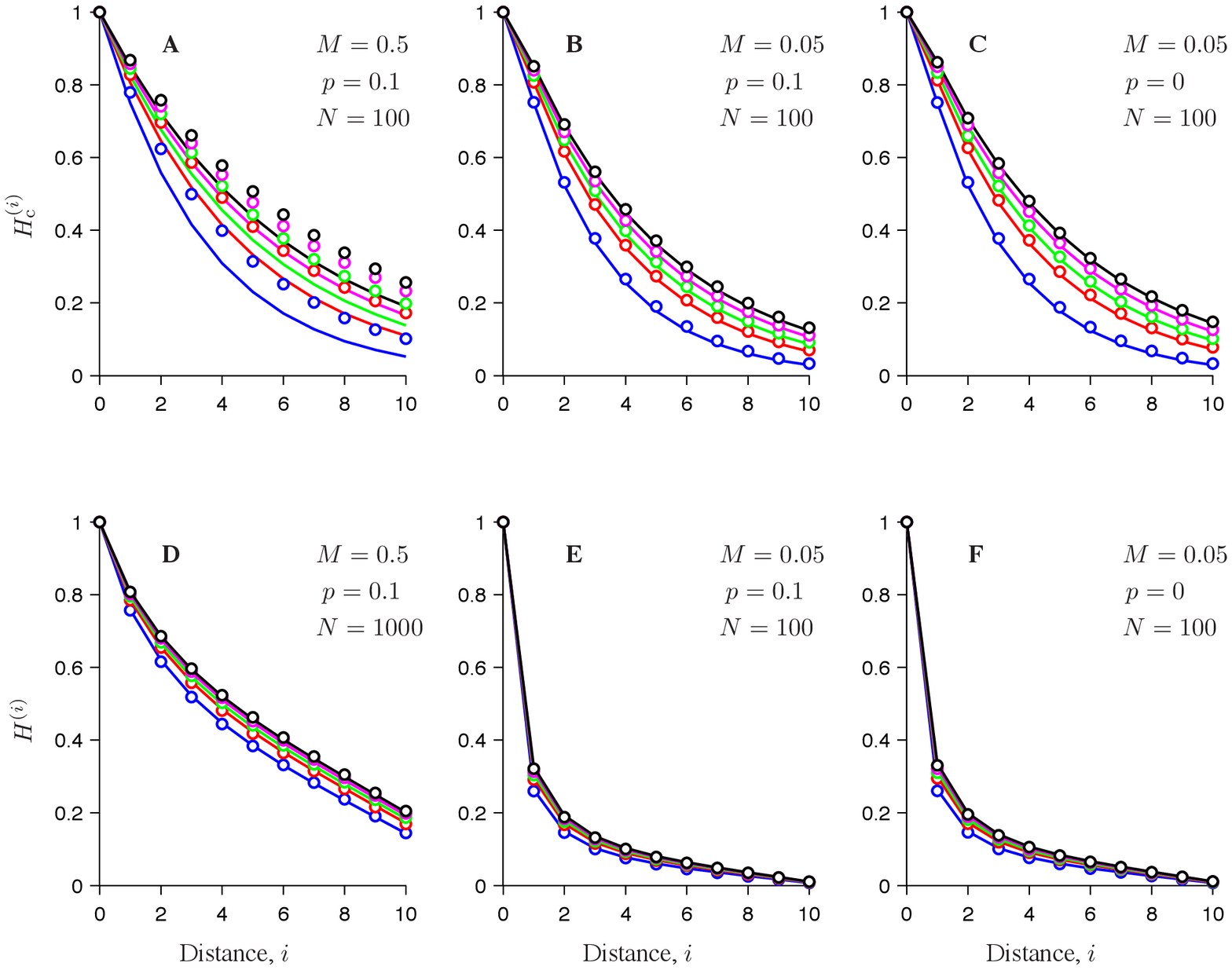}}\hspace*{1cm}
\caption{\label{fig:col_het} Heterozygosity in the colonisation phase ({\bf A}-{\bf C}), and in the equilibrium state ({\bf D}-{\bf F}).  The results corresponding to Eq.~(\ref{eq:h_in}) are shown as solid lines, and the results of computer simulations are shown as symbols. The number of available mates is $s=1$ (blue), $s=2$ (red), $s=3$ (green), $s=5$ (magenta), and $s=10$ (black).  Averages are over $\tau=10^4$ independent realisations of the process of colonisation of empty  islands in {\bf A}, that is over $\tau=2\cdot10^4$ realizations in {\bf B}, and {\bf C}. In panel {\bf D}, averaging is done over $\tau=1.5\cdot 10^7$ generations (the initial $\tau=10^7$ generations being discarded), and over three independent realisations of our model. In panel {\bf E}, averages are over $\tau=4\cdot10^7$ generations (the initial $\tau=5.5\cdot10^7$ generations being discarded) and over four independent realisations of our model. In panel {\bf F}, averaging is done over $\tau=5\cdot 10^7$ generations (the initial $\tau=5\cdot10^6$ generations being discarded) and over five independent realisations of our model. Remaining parameters used: mainland heterozygosity $H^{(0)}=1$, number of islands $k=10$.}
\end{figure}

The heterozygosity in the colonisation phase for different parameters of our model is shown in Fig.~\ref{fig:col_het}{\bf A}-{\bf C}. The analytical result, Eq.~(\ref{eq:h_in}), is shown as solid lines, and the results of our computer simulations are shown as symbols (the solid lines in Fig.~\ref{fig:col_het}{\bf B} correspond to the solid lines in Fig.~4{\bf A} in the main text). We see that the agreement between Eq.~(\ref{eq:h_in}) and the results of computer simulations is good for $M=0.05$, whereas for $M=0.5$, Eq.~(\ref{eq:h_in})  underestimates the results of computer simulations. This is discussed in Section~III in the main text.

\section{Heterozygosity in the equilibrium state}\label{sec:steady_het}
In this section, the equilibrium state within the model introduced in Section~II in the main text is analysed.  Expressions for the equilibrium-state heterozygosity at distance $i=1,\ldots,k$ from the mainland are derived under the assumption that all islands are populated (that is, the colonisation phase is over). 
As discussed in Section~\ref{sec:eff_size} of this supplementary material, the inbreeding coefficient in generation $\tau$ at island $i$, $\epsilon_\tau^{(i)}$, and the coancestry  $\chi_\tau^{(i)}$ contribute to the homozygosity $F_\tau^{(i)}$ at island $i$ in generation $\tau$. The coancestry between islands $i$, and $j$ is equal to the inter-island homozygosity, and  for this case we use the notation  $\chi_\tau^{(i,j)}\equiv F_\tau^{(i,j)}$ ($i\neq j$). 

As mentioned in Section~II of this supplementary material, we assume that the mainland is the only source of genetic variation.
All habitats are assumed to have equal numbers of males and females. The population size on the islands $i=1,\ldots,k$ is assumed to be large (and equal to $2N$), but much smaller than that of the mainland. As before, the heterozygosity on the mainland in generation $\tau=0$ is denoted by $H_{\rm c}^{(0)}$. 

According to the spatial model introduced in  Section~II of the main text, the female-migration rate per island per generation is $2M$ for islands $i=1,\ldots,k-1$, whereas for the mainland and for the island furthest  from the mainland it is equal to $M$. Since the population size on the mainland is much larger than that of a populated island, the process of migration does not affect genetic variation on the mainland. Therefore, we have
$H_\tau^{(0)}=H_{\rm c}^{(0)}$.
For the island populations, we consider separately the inter-island and the intra-island homozygosity. First, we treat sampling from two distinct islands, $i\neq j$. Second, we consider the case of sampling within a single island $i=1,\ldots,k$. Our calculations given below are based on the approach employed in \cite{Wright:1931}.

The inter-island homozygosity $F_{\tau+1}^{(i,j)}$ for $i=0$, $0<j\le k$ satisfies the following recursion:

\begin{align}
F_{\tau+1}^{(0,j)}&=(1-m+\delta_{j,k}\frac{m}{2})F_\tau^{(0,j)}+\frac{m}{2}\left(F_\tau^{(0,j-1)}+(1-\delta_{j,k})F_\tau^{(0,j+1)}\right)\,\,.\label{eq:rec1}
\end{align}
Here $m=2M/N\ll 1$ is the migration rate per island per female per generation, and $\delta_{j,k}$ is equal to unity when $j=k$, and it is zero otherwise.
The inter-island homozygosity for $0<i< k$, $0<j< k$, $i\neq j$, obeys:

\begin{align}
F_{\tau+1}^{(i,j)}&=(1-m)\Bigg[(1-m) F_\tau^{(i,j)}+m\chi_\tau^{(i)}
\left(\delta_{i,j-1}+\delta_{i,j+1}\right)\nonumber\\
&+\frac{m}{2}(1-\delta_{i,j-1})\left(F_\tau^{(i,j-1)}+F_\tau^{(i+1,j)}\right)\nonumber\\
&+\frac{m}{2}(1-\delta_{i,j+1})\left(F_\tau^{(i,j+1)}+F_\tau^{(i-1,j)}\right)\Bigg]+O(m^2)\,\,.
\end{align}
Lastly, when $i=k$, $0<j<k$, we find

\begin{eqnarray}
F_{\tau+1}^{(k,j)}&=&
\big(1-{m\over 2}\big)(1-m)\, F_\tau^{(k,j)}\nonumber\\
&+&\frac{m}{2} (1-\frac{m}{2})  \Big(F_\tau^{(k,j-1)}(1-\delta_{k,j-1})+\chi_\tau^{(k,j-1)}\delta_{k,j-1}\Big)\nonumber\\
&+&\frac{m}{2} (1-\frac{m}{2})\Big(F_\tau^{(k,j+1)}(1-\delta_{k,j+1})+\delta_{k,j+1}\chi_\tau^{(k,j+1)}\Big)\nonumber\\
&+&\frac{m}{2}(1-m)\Big(F_\tau^{(k-1,j)}(1-\delta_{k,j+1})+\chi_\tau^{(k-1,j)}\delta_{k,j+1}\Big)+O(m^2)\,.
\end{eqnarray}
The inbreeding coefficient of the population on island $0<i<k$ satisfies:

\begin{equation}
\epsilon_{\tau+1}^{(i)}=\left(1-m\right)\chi_\tau^{(i)}+\frac{m}{2}\chi_\tau^{(i-1)}+\frac{m}{2}\chi_\tau^{(i+1)}\,\,,
\end{equation} 
and the coancestry is given by:

\begin{align}
\label{eq:rec_chi}
\chi_{\tau+1}^{(i)}&=(1-m)^2\frac{1}{N(1-m)}\left(\frac{1+\epsilon_\tau^{(i)}}{8}(1+\kappa)+\frac{1-\kappa}{4}\chi_\tau^{(i)}+\frac{\chi_\tau^{(i)}}{2}\right)\nonumber\\
&+(1-m)^2(1-\frac{1}{N(1-m)})\left(\frac{3}{4}\chi_\tau^{(i)}+\frac{1+\epsilon_\tau^{(i)}}{8N}+(1-\frac{1}{N})\frac{\chi_\tau^{(i)}}{4}\right)\nonumber\\ 
&+m(1-m)\left( F_\tau^{(i,i+1)} +F_\tau^{(i,i-1)}\right)+O(m^2)\,\,.
\end{align}
For the island furthest from the mainland ($i=k$), we have:

\begin{equation}
\epsilon_{\tau+1}^{(k)}=\left(1-\frac{m}{2}\right)\chi_\tau^{(k)}+\frac{m}{2}\chi_\tau^{(k-1)}\,\,,
\end{equation}

\begin{align}
\chi_{\tau+1}^{(k)}&=(1-\frac{m}{2})^2 \frac{1}{N(1-\frac{m}{2})}\left(\frac{1+\epsilon_\tau^{(k)}}{8}(1+\kappa)+\frac{1-\kappa}{4}\chi_\tau^{(k)}+\frac{\chi_\tau^{(k)}}{2}\right)\nonumber\\
&+(1-\frac{m}{2})^2(1-\frac{1}{N(1-\frac{m}{2})})\left(\frac{3}{4}\chi_\tau^{(k)}+\frac{1+\epsilon_\tau^{(k)}}{8N}+(1-\frac{1}{N})\frac{\chi_\tau^{(k)}}{4}\right)
\nonumber\\
&+m(1-\frac{m}{2}) F_\tau^{(i,i-1)}+O(m^2)\,\,.\label{eq:rec_last}
\end{align}
In what follows we keep only the leading order terms in Eqs.~(\ref{eq:rec1})-(\ref{eq:rec_last}). Moreover, we use the scaled time $t$, where $\tau=\lfloor2N_{\rm e}t\rfloor$, and  $N_{\rm e}$ is the effective population size of an island population under our mating model (see Eq.~(\ref{eq:Ne})). In these units of time, we denote the homozygosity at time $t$ at island $i$ by $F^{(i)}(t)$. 

We finish this section by giving differential equations (with only the leading order terms) for the mainland-island homozygosity, then for the inter-island homozygosity and, lastly, for the intra-island homozygosity. 

For the mainland-island homozygosity we find:

\begin{align}
0=&-\partial_tF^{(0,i)}(t)+M_{\rm e}\left(F^{(0,i+1)}(t)+F^{(0,i-1)}(t)-2F^{(0,i)}(t)\right)\,\,.\label{eq:s1}
\end{align}
Here, $M_{\rm e}=2 M N_{\rm e}/N$, and it is assumed that $i<k$. For $i=k$, we have:

\begin{equation} 
0=-\partial_t F^{(0,k)}(t)+M_{\rm e}\left(F^{(0,k-1)}(t)-F^{(0,k)}(t)\right)\,\,.\label{eq:mig0}
\end{equation}
For the inter-island homozygosity between islands $i$ and $j$, where  $0<i<k$, $0<j<k$, $j\neq i$, we find: 

\begin{align}
0=&-\partial_t F^{(i,j)}(t)+M_{\rm e}\left(1-\delta_{i-1,j}\right)\left(F^{(i,j+1)}(t)+F^{(i-1,j)}(t)\right)\nonumber\\
&+M_{\rm e}\delta_{i-1,j}\left(F^{(i-1)}(t)+F^{(i-1)}(t)\right)+M_{\rm e}\left(1-\delta_{i+1,j}\right)\left(F^{(i,j-1)}(t)+F^{(i+1,j)}(t)\right)\nonumber\\
&+M_{\rm e}\delta_{i+1,j}\left(F^{(i)}(t)+F^{(i+1)}(t)\right)-4M_{\rm e}F^{(i,j)}(t)\,\,.
\end{align}
For $i=k$, $0<j<k$, we obtain:

\begin{equation}
0=-\partial_t F^{(k,j)}(t)+M_{\rm e}\left(F^{(k-1,j)}(t)-2F^{(k,j)}(t)\right)\,\,.\label{eq:mig1}
\end{equation}
Finally, for the homozygosity at distance $0<i<k$ from the mainland we find:

\begin{align}
0=&\left(-\partial_t -1\right){F^{(i)}(t)}+2M_{\rm e}\left(F^{(i+1,i)}(t)+F^{(i-1,i)}(t)-F^{(i)}(t)\right)+1\,\,,
\end{align}
For the island furthest from the mainland, $i=k$, the corresponding expression is:

\begin{equation}
0=\left(-\partial_t -1\right){F^{(k)}(t)}+M_{\rm e}\left(F^{(k,k-1)}(t)-2F^{(k)}(t)\right)+1\,\,.\label{eq:s_last}
\end{equation}
By setting in Eqs.~(\ref{eq:s1})-(\ref{eq:s_last}) the terms involving the time derivative to zero, one finds the expressions for the equilibrium-state homozygosity of the system. The equilibrium-state heterozygosity is obtained upon subtracting the equilibrium-state homozygosity from unity. Upon setting $H^{(0)}=1$, the results shown in Fig.~\ref{fig:col_het}{\bf D}-{\bf F} are obtained. We note that the lines in Fig.~\ref{fig:col_het}{\bf E} correspond to the lines shown in Fig.~4{\bf B} in the main text.

\section{Durations of low- and high-heterozygosity phases}\label{sec:scheme}

\begin{figure}[tp]
\centerline{
\includegraphics[angle=0,width=11.5cm]{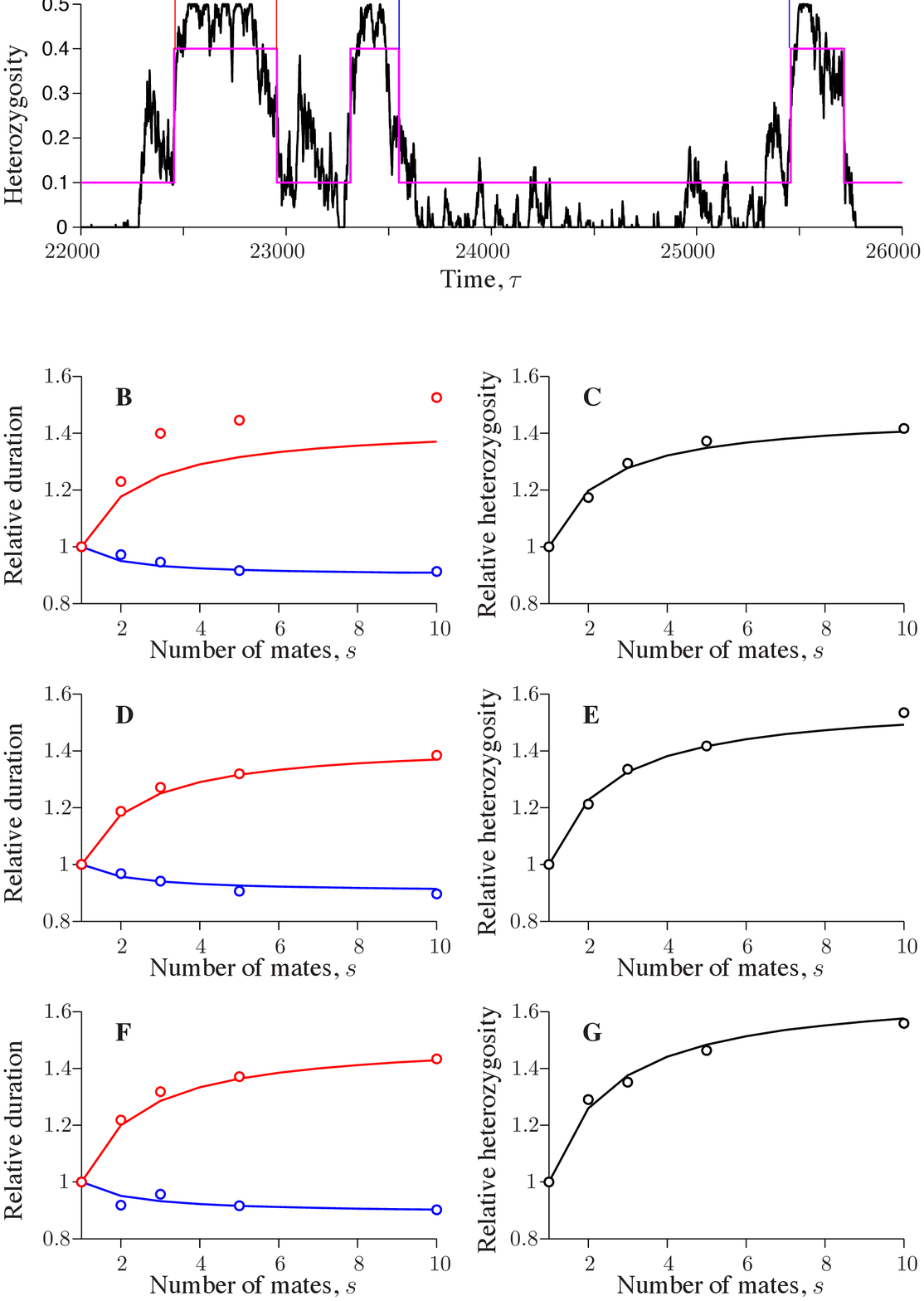}}\hspace*{1cm}
\caption{\label{fig:scheme} ({\bf A}) Illustration of the method used to determine the duration of low- and high-heterozygosity phases. The heterozygosity represented in terms of the low and high phases is shown by the magenta line. The black line depicts the result of computer simulation. The data shown correspond to those in Fig.~5{\bf A} in the main text, for island $10$. ({\bf B}), ({\bf D}), and ({\bf F}) Durations of low- and high heterozygosity phases relative to their corresponding values for $s=1$ (blue and red, respectively). ({\bf C}), ({\bf E}), and ({\bf G}) Equilibrium-state heterozygosity at island $10$ relative to its corresponding value for $s=1$. The results of computer simulations are shown as symbols, and the analytical results are shown as solid lines. The parameters in {\bf B}, and {\bf C} correspond to those in Fig.~\ref{fig:col_het}{\bf D}. The parameters in {\bf D}, and {\bf E} correspond to those in  Fig.~\ref{fig:col_het}{\bf E}.  The parameters in {\bf F}, and {\bf G} correspond to those in  Fig.~\ref{fig:col_het}{\bf F}.}
\end{figure}

A scheme for computing the durations of low- and high-heterozygosity phases is shown in Fig.~\ref{fig:scheme}{\bf A}. As indicated in this figure, we consider values of the  heterozygosity smaller than $0.1$ to be low, and values of the heterozygosity larger than $0.4$ to be high. The maximum value for the low phase ($0.1$) is chosen because a locus is commonly considered monomorphic if the heterozygosity at this locus  is $\le 0.1$ (see  \cite{HartlClark}). The minimum value for the high phase ($0.4$) is chosen since the typical maximum value that the heterozygosity has at the island furthest from the mainland is $\approx 0.5$ for the parameters chosen in Fig.~\ref{fig:scheme}{\bf A}. Note that the maximum value of the heterozygosity at a locus with only $2$ allelic types is equal to $0.5$.

 The method used to record the times of transitions between low- and high-heterozygosity phases in computer simulations is explained next. Say the heterozygosity is in the high phase at time $\tau=0$. We record first the nearest point in time when the heterozygosity becomes less than $0.1$. Say this occurs in generation $\tau_0$. Second, we search for the last  generation before $\tau_0$ in which the heterozygosity has a value larger than or equal to $0.25$ (the middle point between $0.1$ and $0.4$). Say this happens in generation $\tau_1<\tau_0$. We take $\tau_1$ to be the time of transition from the high- to the low-heterozygosity  phase. The  transitions from the low to the high phase are recorded using a similar scheme. 
The durations of the high- and low-heterozygosity phase, $T_{\rm high}$ and $T_{\rm low}$, relative to their corresponding values for $s=1$, computed as explained above, are shown as symbols  in Fig.~\ref{fig:scheme}{\bf B}, {\bf D}, and {\bf F}. For comparison, we show in Fig.~\ref{fig:scheme}{\bf C}, {\bf E}, and {\bf G} the corresponding equilibrium-state heterozygosities relative to their values for $s=1$.

Next, we briefly explain how to estimate $T_{\rm low}$, and $T_{\rm high}$ analytically. The heterozygosity can  switch from the low to the high phase if new genetic material comes to the population, and if this material is not lost due to random genetic drift. Recall that, in our model, migration is the only process bringing new genetic material to islands. However, some migrations bring genetic material that already exists in a given island population (but note that when $H^{(0)}=1$, then the first island receives new genetic material with each migration). Yet, it is possible to estimate the effective successful migration rate per allele per generation, $m_{\rm e}^{(i)}$, at a given distance $i$ from the mainland using the analytical results derived for the equilibrium-state heterozygosity at island $i$, namely $H^{(i)}$. Here, the term \lq successful\rq ~implies that migrants bring new genetic material to the population. Note that under the assumption that a new allelic type is introduced to the population at distance $i$ from the mainland at rate $m_{\rm e}^{(i)}$ per allele per generation, the equilibrium-state heterozygosity is computed as  

\bee
H^{(i)}=\frac{4m_{\rm e}^{(i)}N_{\rm e}}{1+4m_{\rm e}^{(i)}N_{\rm e}}\,\,,
\ee
where $2m_{\rm e}^{(i)}N_{\rm e}$ is the total number of new allelic types introduced in the population per generation (this expression is analogous to the usual expression for the heterozygosity which involves the scaled mutation rate, see Eq.~(\ref{eq:F})).
In the case $m_{\rm e}^{(i)}N_{\rm e}\ll 1$, it follows that typically every $(2m_{\rm e}^{(i)}N_{\rm e})^{-1}$ generations, one mother comes to an island carrying a new allele. Furthermore, it is known \cite{Ewens:1982} that a large haploid population with effective size $2N_{\rm e}$ spends on average $2/N_{\rm e}$ generations in the state with this allele having the frequency $N_{\rm e}$ before fixation occurs. It follows that the typical waiting time for such a successful establishment of new genetic material at island $i$ is equal to $(4m_{\rm e}^{(i)})^{-1}$. This estimate agrees well with the results of our computer simulations (results not shown). 

We now explain how to estimate the average duration of the high-heterozygosity phase, $T_{\rm high}$. In the case of rare income of new allelic types to island $i$, $m_{\rm e}^{(i)}N_{\rm e}\ll 1$, the island population at a given locus typically has at most $2$ alleles. Thus, $T_{\rm high}$ can be approximated by the time that a locus with two alleles in a Wright-Fisher population of $N_{\rm e}$ diploid individuals needs to reach fixation. Note that under the condition $m_{\rm e}^{(i)}N_{\rm e}\ll 1$, fixation occurs typically much before new genetic material arrives to the population. The number of generations until fixation in a diploid population at a locus with two alleles, $\tau_{\rm loss}$, is \cite{HartlClark}
 
 \bee\label{eq:alpha}
\tau_{\rm loss}(\alpha)\approx -4N_{\rm e}[\alpha\ln(\alpha)+(1-\alpha)\ln(1-\alpha)]\,\,.
\ee                                                                                                           
Here $\alpha$ is the initial frequency of a given allelic type. For a given value of the number of available mates, $s$, we compute $T_{\rm high}$ upon integrating Eq.~(\ref{eq:alpha}) from $\alpha=0.1$ to $\alpha=0.9$. The integral boundaries correspond to the value of heterozygosity $\approx 0.2$, which is close to the value of $0.25$ in the method used for determining the transition time between the low and the high phase. We note that Eq.~(\ref{eq:alpha}) results in underestimated time of fixation if the rate of income of new genetic material is not too small. 
This is confirmed by the results shown in  Fig.~\ref{fig:scheme}{\bf B}, {\bf D}, and {\bf F}. Note that the solid lines in Fig.~\ref{fig:scheme}{\bf B}, and {\bf C} correspond to the lines in Fig.~5{\bf B}, and {\bf C} (respectively) in the main text. 

